\documentclass[12pt,draftcls,onecolumn]{IEEEtran}
\usepackage{lmodern}
\usepackage{amstext}	
\usepackage{verbatim}   

\usepackage{algorithm}
\usepackage{amssymb}  
\usepackage{amsmath}

\usepackage{graphics}
\usepackage{tikz}
\usepackage{booktabs}
\usepackage{bm}			
\usepackage{graphicx}
\usepackage{caption}
\usepackage{subcaption}
\usepackage{stfloats}   
\usepackage{hyperref}
\usepackage{cite}
\usepackage{gensymb} 

\definecolor{gold}{rgb}{0.85,.66,0}
\definecolor{green2}{rgb}{0.01,0.5,0.01}
\definecolor{cian}{rgb}{.02,.7,.95}
\definecolor{ppp}{rgb}{.7,.3,.82}

\usepackage{soul}  

\begin{document}
\title{Pre-distortion and Pre-equalization for Non-Linearities and Low-Pass Effect Mitigation in OFDM-VLC Systems}

\author{Luis Carlos Mathias , 
	Jose Carlos Marinello Filho
	and Taufik Abrao
	\thanks{L. C. Mathias , J. C. M. Filho
		and T. Abrao are with the Department of Electrical Engineering, State University of Londrina, Rod. Celso Garcia, PR-445 Km 380, CEP 86051-990, Londrina-PR, Brazil (e-mail: luis.mathias@uel.br; 
		zecarlos.ee@gmail.com; 
		taufik@uel.br).}
}

\maketitle

\begin{abstract}
The orthogonal frequency division multiplexing (OFDM) transmission has shown promise in applications of visible light communication (VLC). However, the variation of the nonlinearity of the optical power emitted by the high power light emitting diode (HPLED) as a function of current and temperature implies in drastic OFDM-VLC performance degradation. The first part of this work, experimentally confirms and models this degradation due to temperature in a high power white HPLED. The higher attenuation at high frequencies, which is inherent to the HPLED and which is accentuated by the effect of the intrinsic capacitance of the photodiode, is another factor of degradation due to the reduction of the signal-to-noise ratio (SNR) at the receiver for such frequencies. For the mitigation of these effects, we propose a pre-distortion and digital pre-equalization scheme using a luminous feedback signal in the transmitter module. The system is modeled so that the operating points are mathematically deduced and evaluated by simulations and by an experimental setup. By allowing the linearization of the transmitted light signal and the maintenance of an average SNR in all OFDM subcarriers,  the performance improvement is confirmed in comparison with other schemes, such as with non-predistortion, pre-distortion with fixed parameters, and simple post-equalization.
\end{abstract}


\section{Introduction}\label{sec:intro}

The increasing demand for data transmission, the use of HPLED (High Power Light Emitting Diode) lighting, the scarcity and high cost of the radio frequency (RF) spectrum  have made studies for the application of visible light communication (VLC) prominent \cite{pathak_2015}.
These and other advantages corroborate in placing VLC as one of the key technologies for 5G wireless systems \cite{haas_chen_2017}.

However, the design of the electronic circuit of the VLC transmitter and VLC receiver present several challenges. On the transmitter side, the attenuation of the light signal emitted and the nonlinearity of the optical power in relation to the current supplied to the HPLED degrade the performance of the orthogonal frequency division multiplexing (OFDM) in VLC. On the receiver side, another limitation is the intrinsic capacitance of the junction of the photodiode (PD) that makes it difficult to implement a transimpedance amplifier (TIA) for large band of frequency.

The higher attenuation of the signal transmitted by the high frequencies is directly affected by the construction characteristics of the white HPLED.
This depends on the process of generating white light in HPLED luminaires which is basically divided into two forms \cite{pathak_2015}. In the first form, the white light is obtained from the combination of the blue HPLED light when crossing a yellow layer of the chemical element phosphorus. 
The second form uses the additive mixture of the lights generated by red, green and blue (RGB) colored HPLEDs. The first-mentioned HPLED lamp, although easier to deploy and less costly, has a limited switching speed of up to a few $MHz$, due to the luminous persistence of the phosphor layer \cite{le_minh_2009}.

Another degrading factor are the nonlinearity issues \cite{dimitrov2013}. The first source of nonlinearity occurs between the voltage applied to the HPLED and your current \cite{sheu2017}. In \cite{sterckx_2012} and \cite{fuada_2016}, the nonlinearity of the active components of the modulator was mitigated using  A-class amplifier circuit for analog-front-end (AFE) as current sink with feedback loop. However, it does not combat another nonlinearity problem for modulation in HPLED type transmitters: the nonlinearity between the current applied to the HPLED and the respective optical power obtained \cite{dimitrov2013,2016chapterModelingLED,datasheetLED,led1w}. According to \cite{elgala_2010} and \cite{2016chapterModelingLED}, this nonlinearity can be modeled using a second order polynomial function.

However, for lighting purposes, several studies have verified a strong variation of the optical efficiency, and consequently, of transmitted optical powers, of HPLEDs when faced with factors such as temperature and aging \cite{ yu2017, rao_2012, ChiChou2010, keppens2009evaluation}. Differences between manufacturing batches, etc., which can further aggravate this scenario making even more ineffective the correction of nonlinearities by eventual factory fixed modeling.

The main limitation of the VLC receiver is the capacitance of the photodiode receiver which limits the bandwidth of the TIA. This capacitance is directly proportional to your sensing area \cite{2012_ghassemlooy}. It also depends on the size of the depletion region generated by the polarization of the photodiode. This because the depletion region is a non-electrically conductive region \cite{boylestad2002}, it functions as a dielectric. Thus, increasing the distance between the conducting regions has the same effect as increasing the distance between the parallel conductive plates of a capacitor, i.e., its capacitance decreases.

The work of \cite{le_minh_2009} minimizes the problem of light attenuation in HPLED by applying post-equalization but also does not solve the nonlinearity problem in HPLED emission. The experimental DAC proposed in \cite{yew_2013}, which is composed by an array of LEDs, is able to correct the problem of the HPLED nonlinearity; however, the scheme requires a large number of HPLEDs to obtain a reasonable resolution. This requirement causes yet another problem of nonlinearity due to the different channel attenuations generated as a function of position and distance of the HPLEDs. A drive circuit proposed in \cite{zuhdi_2016} employs on-chip optical feedback technique to suppress the HPLED nonlinearities. However, it does not evaluate the attenuation problem for larger frequencies.

Against this background, the present work proposes a digital pre-distortion (DPD) and a pre-equalization (Pre-Eq) schemes based on the light signal feedback on the transmitter module aiming at mitigating the non-linearities and low-pass effect inherent to OFDM-VLC systems.  In this sense, the DPD function allows the system learn on the HPLED nonlinearity parameters and applies correction to mitigate this effect. Furthermore, the Pre-Eq estimates the attenuation of the OFDM subcarriers and establishes a pre-equalization strategy in order to maintain an average target SNR on each subcarriers signal at the receiver side. Both processes estimate the combined effect between transmitter and receiver and are fully computationally executed in the transmitter, decreasing the complexity of the receiver.
 
The remainder sections of the paper is organized as follows. The VLC system model is presented in Section \ref{sec:vlc}. 
Section \ref{sec:HPLED analyser} describes the developed prototype to caracterize the variation of the nonlinearity of the current against the transmitted optical power in an HPLED as a function of the temperature. Section \ref{sec:proposed_scheme} presents the proposed pre-distortion and pre-compensation schemes for OFDM-VLC to mitigate the nonlinearity and attenuation  problems. Section \ref{sec:results} explores the numerical and experimental results that validate the proposal. Finally, Section \ref{sec:conc} offers the main conclusions.        

\section{VLC System}\label{sec:vlc}
\subsection{VLC Channel Model}\label{subsec:vlc_channel}

Considering that the position and orientation of the HPLED is within the field of view of the PD, the DC optical channel gain between the receiver and the HPLED can be simplified by \cite{barry_1993}:
\begin{equation}
\Omega_{\rm dc}  =  \frac{(n_L + 1)A_{\rm pd}}{{2\pi }}\cdot \frac{\cos^{n_L}(\phi)\cos(\theta)}{R^2};
\label{eq:omega}
\end{equation}
where $\phi$ is the angle between the orientation versor of the HPLED transmitter and the incidence vector, $\theta$ is the angle between the receiver orientation versor and the incidence vector, $R$ is the distance between the HPLED transmitter and the receiver PD, $A_{\rm pd}$ is the area of the PD in $m^2$, and $n_L$ is the mode number of the Lambertian distribution\footnote{The greater the value of $n_L$, the more directive is the distribution of light.}. Thus, considering an optical power $P_T$ transmitted by HPLED, the optical power captured from this signal by the receiver is given by $P_R=\Omega_{\rm dc} P_T$.

\begin{figure}[!htbp] 
\centering\includegraphics[width=.8\textwidth]{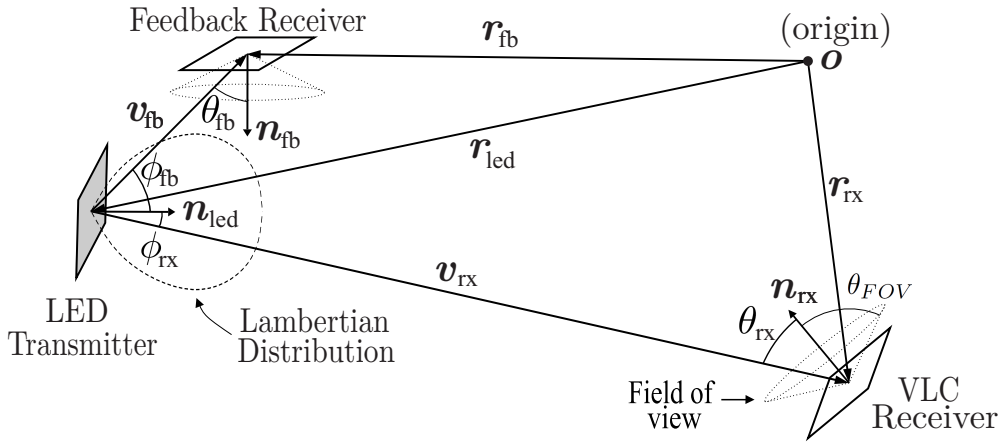}
	\caption{ Schematic diagram of the system geometry.}
	\label{fig:system_geometry} 
\end{figure}

Fig. \ref{fig:system_geometry} shows the schematic diagram of the system geometry proposed in this work. 
Considering identical PDs, and the definition of the internal product among vectors, the second term of \eqref{eq:omega} can be converted to \cite{IEEEhowto:sahin}:
\begin{equation}
\Omega_{\rm{rx}}  = - \frac{{(n_L + 1)A_{\rm pd} }}{{2\pi }} \cdot \frac{{\left( {\bm{v}_{\rm{rx}} ^\intercal \bm{n}_{\rm{led}} } \right)^{n_L} \bm{v}_{\rm rx} ^\intercal \bm{n}_{\rm{rx}} }}{{\left\| {\bm{v}_{\rm{rx}} } \right\|_2^{n_L + 3} }};
\label{eq:omega_rx}
\end{equation}
\begin{equation}
\Omega_{\rm{fb}}  = - \frac{{(n_L + 1)A_{\rm pd} }}{{2\pi }} \cdot \frac{{\left( {\bm{v}_{\rm{fb}} ^\intercal \bm{n}_{\rm{led}} } \right)^{n_L} \bm{v}_{\rm fb} ^\intercal \bm{n}_{\rm{fb}} }}{{\left\| {\bm{v}_{\rm{fb}} } \right\|_2^{n_L + 3} }};
\label{eq:omega_fb}
\end{equation}
where $\bm{r}_{\rm{led}}$, $\bm{r}_{\rm{fb}}$ and $\bm{r}_{\rm{rx}}$ are the position vectors and $\bm{n}_{\rm{led}}$, $\bm{n}_{\rm{fb}}$ and $\bm{n}_{\rm{rx}}$ are the normal vectors of the VLC LED transmitter, of the feedback receiver and of the VLC receiver respectively. All vectors defined in $\mathbb{R}^{3 \times 1}$. Finally, $\bm{v}_{\rm{rx}}=\bm{r}_{\rm{rx}}-\bm{r}_{\rm{led}}$ and 
$\bm{v}_{\rm{fb}}=\bm{r}_{\rm{fb}}-\bm{r}_{\rm{led}}$. 
\subsection{HPLED transmitter and PD receiver}\label{subsec:vlc_tx}
Due to the greater commercial use of HPLED with phosphor layer and its higher selective attenuation characteristic, this work uses this type of HPLED as a worst case  evaluation.
In the signal reception, PDs based on PIN junction are usually employed because it allows easy conversion of the light signal to current up to tens of $MHz$ \cite{pathak_2015}. 
The current generated is proportional to the irradiance captured by the PD sensing area \cite{vlc_multiple_leds}. In this context, the electric gain $G_E$ in [V/A] as a function of frequency $f$ can be given by:
\begin{equation}
G_E(f)= S_{\rm{led}}(f)\Omega_{\rm dc}  {R_{\rm{pd}}(f)} G_{\rm tia}(f),
\label{eq:electric_gain}
\end{equation}
where $S_{\rm{led}}$ is the HPLED conversion factor in [W/A], $R_{\rm{pd}}$ is the PD responsivity generally presented in the data sheets in [A/W], and $G_{\rm tia}$ is the transimpedance amplifier (TIA) gain in [V/A].

\subsection{Noise Model}\label{subsec:noise}

In the VLC receiver, the dominant types of noise are the shot noise generated by the photocurrent, and the thermal noise coming from receiver electronics \cite{2012_ghassemlooy,2016_3-D_RSS}. The noise can be modeled as Gaussian process with zero-mean and variance \cite{2004_komine}:
\begin{equation}
	\sigma^2_n = \sigma^2_{\rm shot} + \sigma^2_{\rm thermal}.
\label{eq:noise}
\end{equation} 
Thus, the noise is added to the PDs photo-current as an additive white Gaussian noise (AWGN). This noise is shaped by the transfer function of the preamplifier topology. In this work, it will be considered a receiver with photo-detector with PIN PD and TIA with field effect transistor (FET) \cite{receiver_design}. The major source of noise in the optical link is the photo-generated shot noise which corresponds to the fluctuations in the count of the photons collected by the receiver \cite{2004_komine}. Its variance can be determined by:
\begin{equation} 
\sigma _{\rm shot} ^2 = 2qB \left[ R_{\rm pd} \left( P_{R} + A_{ \rm pd}p_{\rm bs}\Delta \lambda \right) + I_{\rm dc} \right],
\end{equation} 
where $q$ is the elementary charge, $B$ is the TIA bandwidth in Hz,  $p_{\rm bs}$ is the background spectral irradiance, $\Delta\lambda$ is the bandwidth of the optical filter and $I_{\rm dc}$ is the dark current.

Thermal noise is independent of the received optical signal and can be determined in terms of noise in the feedback resistor and noise in the FET channel. Each term contributes to the thermal noise variance:
\begin{equation} 
\sigma _{\rm thermal} ^2 = \frac{{8\pi k_B T_K }}{{G_{\rm ol} }}C_{\rm pd} A_{\rm pd}I_2 B^2 + \frac{{16\pi ^2 k_B T_K \Gamma }}{{g_m }}C_{\rm pd}^2 A_{\rm pd}^2 I_3 B^3;
\end{equation} 
where $k_B$ is the Boltzmann's constant, $T_K$ is the absolute temperature, $G_{\rm ol}$ is the open loop gain, $C_{\rm pd}$ is the capacitance per unit area of the PD, $\Gamma$ is the FET channel noise factor, $g_m$ is the FET transconductance, $I_2 = 0.562$ is the TIA bandwidth factor, and $I_3 = 0.0868$ is the TIA noise factor \cite{2004_komine,receiver_design}.

\section{Development of an HPLED analyzer}\label{sec:HPLED analyser}

\begin{figure}[!htbp]
\centering\includegraphics[width=0.7\linewidth]{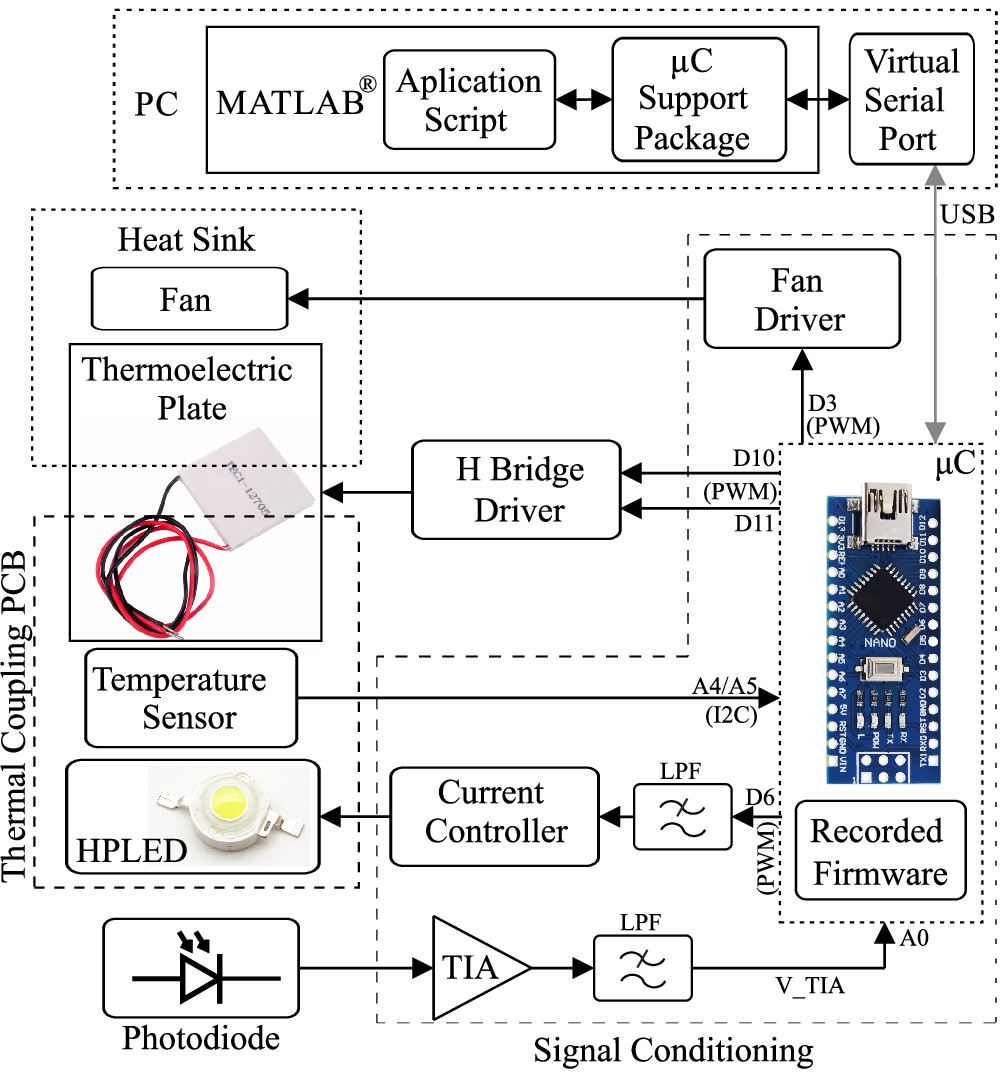}
	\\ \vspace{-.1cm} a) \\ \vspace{.2cm}
\includegraphics[width=0.75\linewidth]{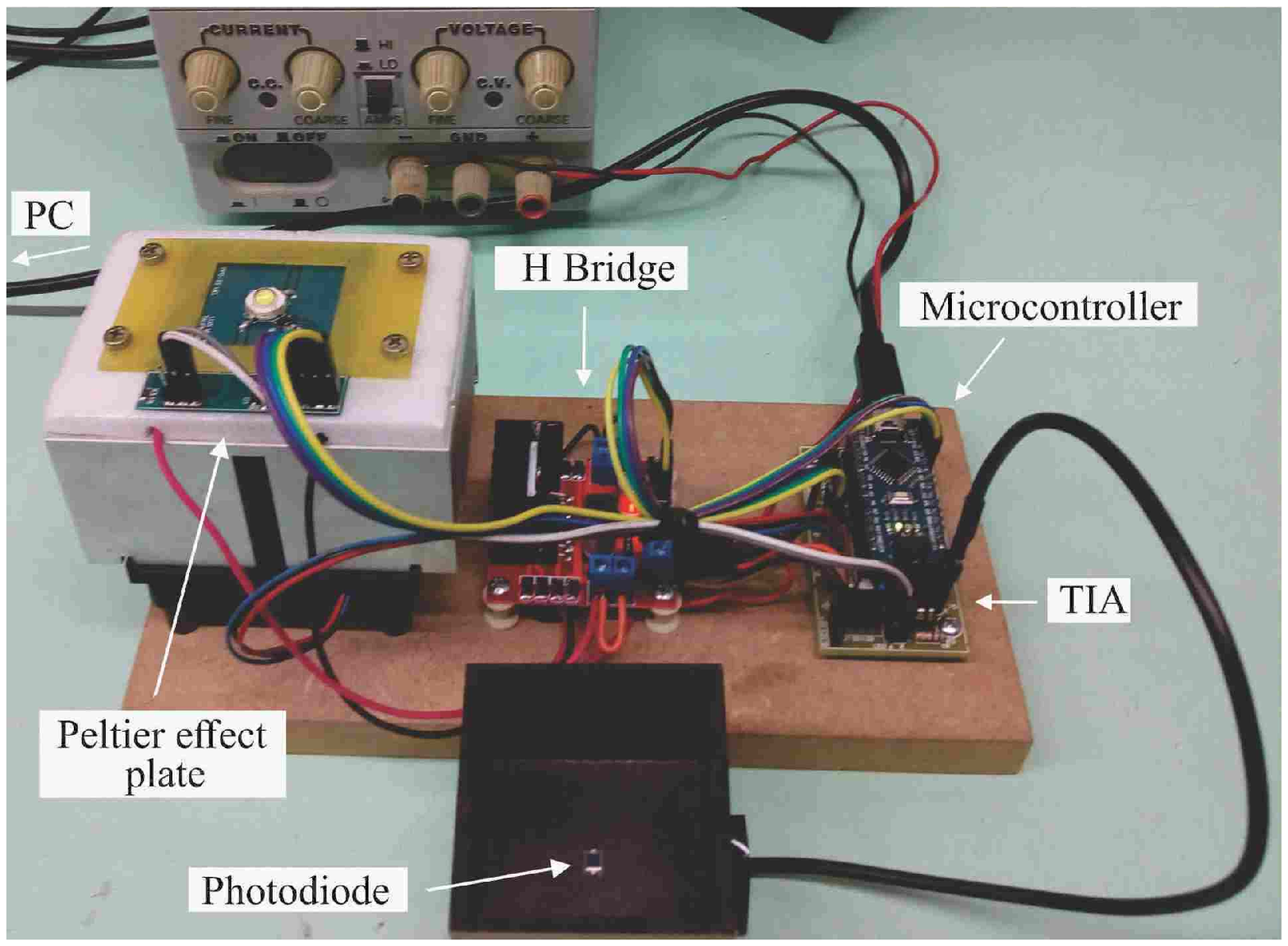}
	\\ b)
	\caption{Block diagram in a) and photo in b) of the experimental arrangement to verify nonlinearity as a function of current and temperature of the HPLED.}
	\label{fig:arranjoanalisalinearidadetemperatura}
\end{figure}

\begin{figure}[!htbp]
\centering\includegraphics[width=0.4\linewidth]{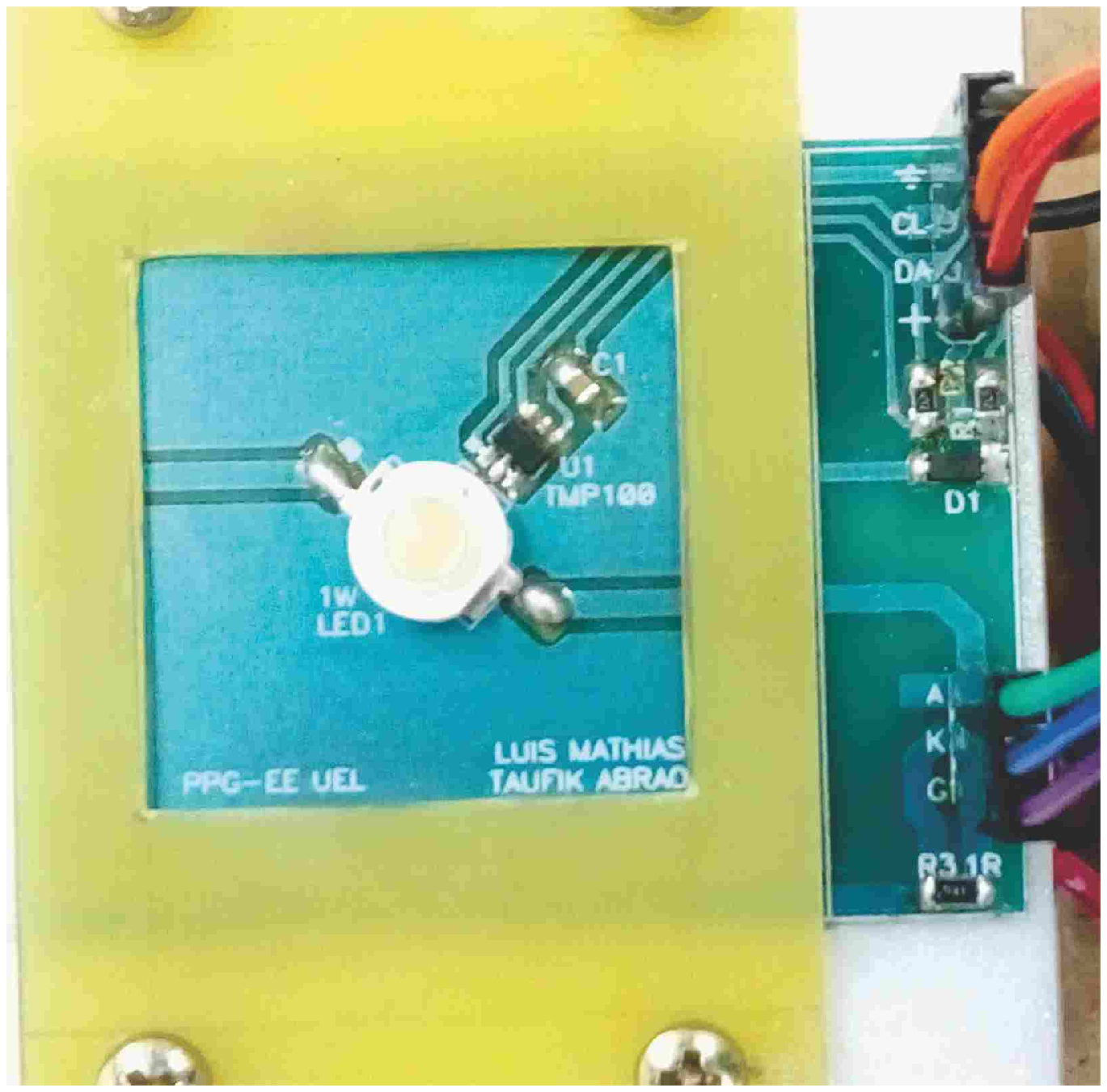}
	\caption{Details of the thermal coupling plate. }
	\label{fig:ledsensor}
\end{figure}

The prototype depicted in Fig. \ref{fig:arranjoanalisalinearidadetemperatura} was developed with the purpose of capturing the behavior of the luminous flux as a function of the current and also of the temperature of a HPLED.
Fig. \ref{fig:ledsensor} shows the detail of the thermal coupling printed circuit board (PCB) between aluminum heat sink with fan, Peltier effect plate, HPLED and temperature sensor. The sensor used was the TMP100 which is calibrated at the factory with typical accuracy of $\pm 1 \degree$C \cite{tmp100}.

The procedure for extracting the experimental data consisted of initially keeping the HPLED off, controlling the analysis temperature of the HPLED, then triggering the analysis current of the HPLED for a very short period of time. The short operating time is intended not to significantly change the temperature of the HPLED junction. The luminous intensity of the emitted pulse is then captured by the analog to digital converter (ADC) through the voltage signal generated by the TIA plus PD arrangement.
The PD BPW34 from the manufacturer Osram and the white HPLED 1W from the manufacturer Multicomp, both aligned with each other at a distance of 3.9 cm, were used. The TIA gain was adjusted by $32\,k\Omega$, i.e., $V_{ \rm{tia}}= 32k\Omega\cdot I_{\rm pd}$.

\begin{figure}[!htbp] 
	\centering \includegraphics[width=.68\textwidth]{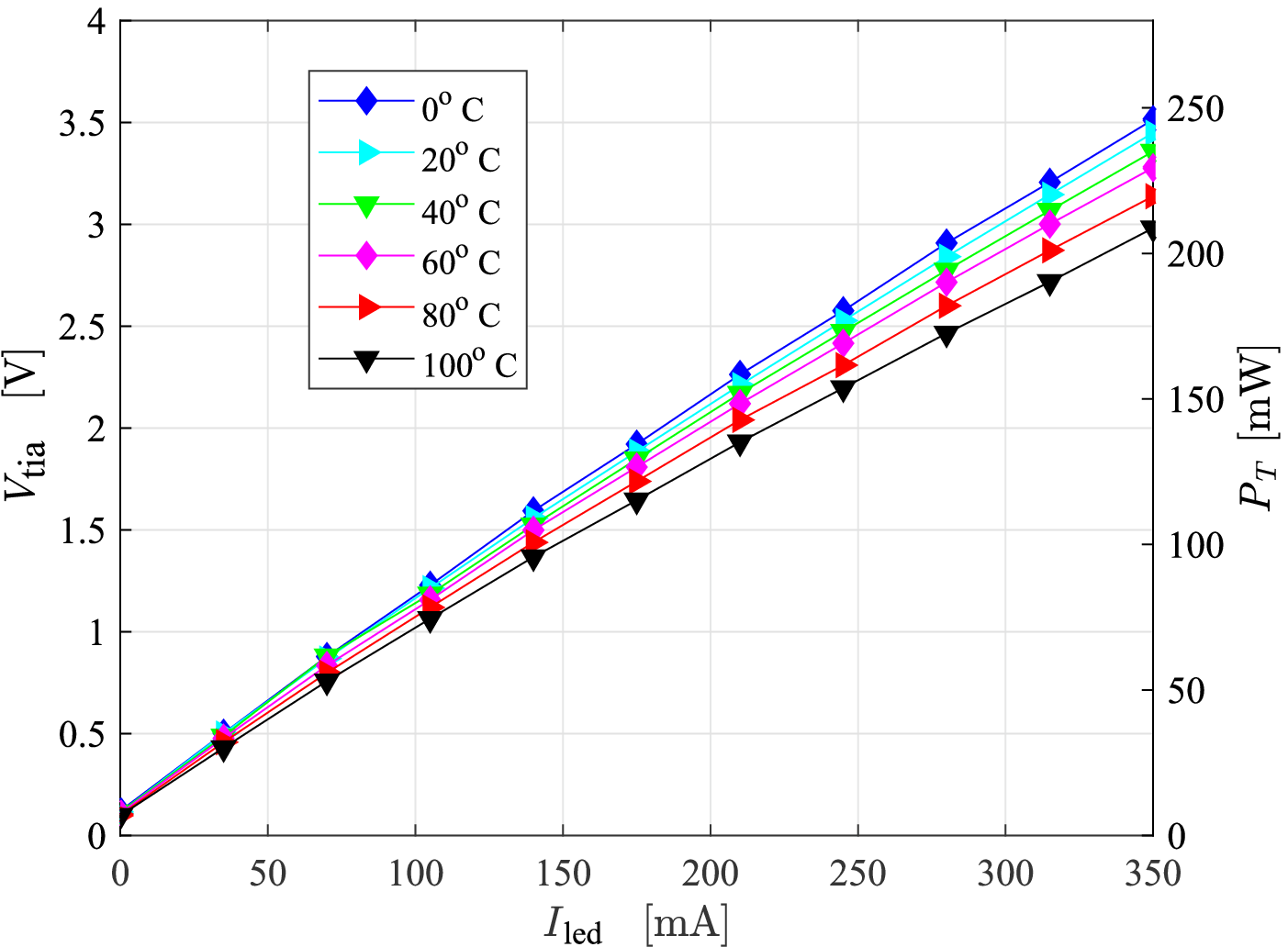}
	\caption{
		Experimental verification of the variation of nonlinearity of a Multicomp 1W white HPLED.}
	\label{fig:ajuste_led_nao_linear} 
\end{figure}

Fig. \ref{fig:ajuste_led_nao_linear} shows the TIA output signal as a function of the DC current applied to the HPLED. In addition to confirming the nonlinearity of the voltage generated by the TIA as a function of the current in the HPLED, it is possible to verify that the luminous efficiency decreases with the increase of the temperature in the HPLED.

The $V_{\rm tia}$, the $I_{\rm pd}$, the irradiance captured by the PD and the luminous flux $\phi_v$ emitted by the HPLED are directly proportional to each other. 
Considering the datasheet information that with $I_{\rm led}^0 =$ 350 mA and $T^0 = 30\degree C$, the HPLED emits $\phi_v^0$ = 115 lumens \cite{led1w}. Indeed, TIA output signal measured experimentally under such conditions was $V_{\rm tia}^0 =$ 3.457V. The luminous flux $\phi_v$ for a LED current and a temperature $T$ can be estimated by:
\begin{equation}
\phi_v(I_{\rm{led}},T) = \phi_v^0 \frac{V_{\rm{tia}}(I_{\rm{led}},T)}{V_{\rm{tia}}^0} =\frac{115}{3.457} V_{\rm{tia}} (I_{\rm{led}},T).\
\end{equation}

The conversion from luminous flux $\phi_v$ to the transmitted optic power $P_T$ can be realized 
for phosphor-coated blue LED  with $P_T=2.1 \cdot \phi_v$ in $[\rm{mW/lm}]$ \cite{2008broadbandLED}. Thus, the right y-axis in Fig. \ref{fig:ajuste_led_nao_linear} reveals a scale resizing the experimental data denoting $P_T$. According to Subsection \ref{subsec:vlc_tx}, the $P_T$ curve can be adjusted by a polynomial function:
\begin{equation}
	P_T =  A \cdot I_{\rm{led}}^2 + B \cdot I_{\rm{led}} + C.
\end{equation}
Thus, for all temperatures analyzed\footnote{The rationale for analyzing higher temperatures is due to the use in tropical environments or in cases of undersizing of heat sinks.}, the parameters of the polynomial fit were recorded in Table \ref{tab:fit_parameters}. All the adjustments have resulted in excellent correlation coefficients $R$. Thus, the variation of the nonlinearity as a function of the temperature has been confirmed.
\begin{table}[htbp]
\centering
\caption{Parameters of polynomial adjustments}
\begin{tabular}{c|ccc|c}
\hline
 $T$     & \multicolumn{4}{c}{Parameters} \\
$[\degree C]$     & $A$     & $B$     & $C$    & $R$ \\
\hline
-10   & -0,23735 & 0,764263 & 0,009285 & 0.99997 \\
0     & -0.24969 & 0.763086 & 0.008686 & 0.99998 \\
10    & -0.22750 & 0.751208 & 0.008882 & 0.99999 \\
20    & -0.24060 & 0.745831 & 0.008781 & 0.99998 \\
30    & -0.22566 & 0.742316 & 0.008848 & 0.99997 \\
40    & -0.25976 & 0.734123 & 0.008910 & 0.99993 \\
50    & -0.26767 & 0.730652 & 0.008128 & 0.99997 \\
60    & -0.26323 & 0.722152 & 0.008205 & 0.99998 \\
70    & -0.27341 & 0.713003 & 0.007769 & 0.99997 \\
80    & -0.27189 & 0.698047 & 0.007694 & 0.99997 \\
90    & -0.26485 & 0.680774 & 0.007487 & 0.99997 \\
100   & -0.25532 & 0.661240 & 0.007331 & 0.99997 \\
		\hline
	\end{tabular}%
	\label{tab:fit_parameters}%
\end{table}%

\section{Proposed Pre-distortion and Pre-equalization OFDM Scheme}\label{sec:proposed_scheme}
\begin{figure}[!htbp]
\centering\includegraphics[width=.95\textwidth]{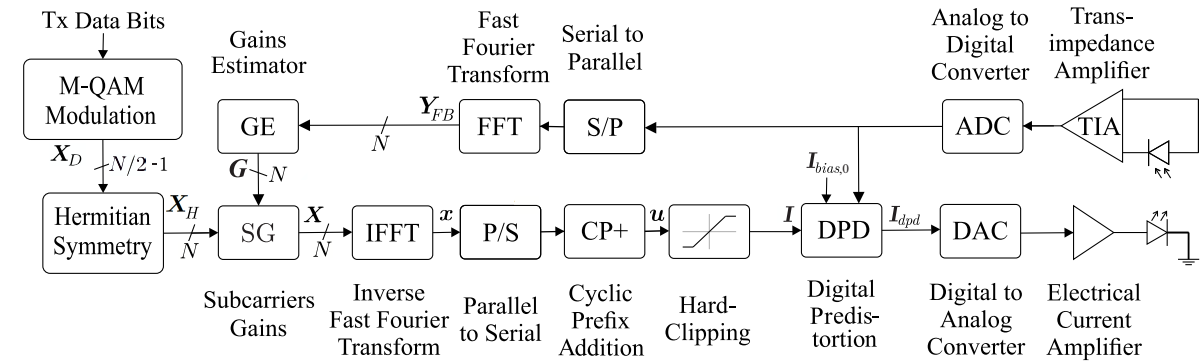}\\ 
 \vspace{.1cm} a) VLC Transmitter \vspace{.2cm}\\
 \includegraphics[width=.97\textwidth]{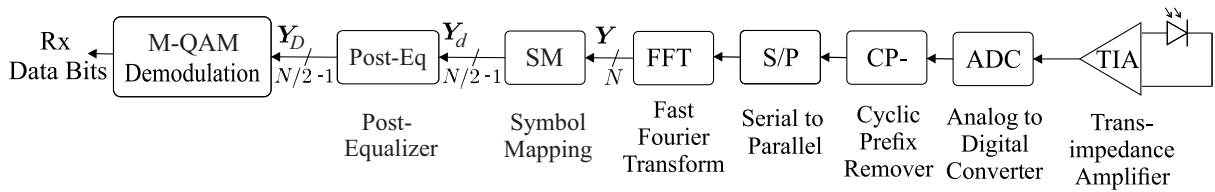}\\
 b) VLC Receiver\\
 \caption{Proposed OFDM Scheme for HPLED Linearization and Attenuation compensation.}
	\label{fig:proposed_scheme} 
\end{figure}

The proposed architecture for the non-linearity compensation of the HPLED and also considering pre-equalization of the powers of the subcarriers detected in the receiver is depicted in Fig. \ref{fig:proposed_scheme}. It is used an {\it intensity modulation / direct detection} (IM/DD) scheme with DC-biased optical OFDM (DCO-OFDM).

\subsection{DCO-OFDM VLC transmitter} \label{subsec:dco_ofdm_vlc_tx}
The data bits are M-QAM modulated generating the vector of symbols $\bm{X}_D$. Considering a length $N$ of the input vector $\bm{X}$ of the inverse fast Fourier transform (IFFT), the length of $\bm{X}_D$ is $N/2-1$ because of the Hermitian symmetry and of the element 
responsible for the DC level is null; hence, not interfering with the bias added by digital pre-distortion (DPD) to the OFDM frame in order to keep it positive.

With the purpose in obtaining purely real-time points at the out of the IFFT block, the vector $\bm {X}_H$ of size $N$ is generated after applying the Hermitian symmetry:
\begin{equation}
\bm{X}_H[n] = \left\{ {\begin{array}{*{20}c}
	{0} \hfill & {;\,n = 0, N/2} \hfill  \\
	{\bm{X}_{D} [n]} \hfill & {;\,n = 1, \ldots , N/2-1 } \hfill  \\
	{\bm{X}_{D} ^* [N - n]} \hfill & {;\,n = N/2+1, \ldots ,N - 1.} \hfill  \\
	\end{array}} \right.
\label{eq:XH}
\end{equation}
After that, the powers of the OFDM subcarriers are modified in the subcarriers gain (SG) block  by:
\begin{equation}
\bm{X} = \bm{X}_H \circ \bm{G},
\label{eq:gain_multiplication}
\end{equation}
where $\circ$ is the element-wise product operator, and $\bm{G}$  is the gain vector obtained by the gain estimator (GE) block. Details on the GE estimation are discussed in Subsection \ref{subsubsec:ge_flat} and \ref{subsubsec:ge_non_flat}.
The IFFT operation is performed in the vector $\bm{X}$ to generate the signal in the time domain.
This vector is converted from parallel to serial (P/S) and appended the cyclic prefix (CP) with length $N_{\rm{CP}}$, obtaining the OFDM frame in signal $\bm{u}[i]$ for $0 \le i \le (N +N_{\rm{CP}} - 1)$. 
Then, the signal is hard-clipped aiming to fit the dynamic range of the driver:
\begin{equation}
\bm{I} [i] = \left\{ {\begin{array}{*{20}c}
	{I_u } \hfill & {;\,\bm{u} [i] > I_u } \hfill  \\
	{\bm{u} [i]} \hfill & {;\,I_l  \le \bm{u} [ i] \le I_u } \hfill  \\
	{I_l } \hfill & {;\,\bm{u} [i] < I_l; } \hfill  \\
	\end{array}} \right.
\end{equation}
where $I_u$ and $I_l$ are the upper and the lower current limits of modulation, respectively.

The severity of clipping suffered by a signal is quantified by the clipping factor that is defined as the number of standard deviations per half of the dynamic range \cite{clipping_noise_ofdm_owc}:
\begin{equation}
\gamma  = { \frac{{I_u  - I_l }}{{2\cdot \sigma_x }}}.
\label{eq:Fator_Clipping}
\end{equation}

\subsection{Digital Pre-distortion block} \label{subsec:dpd}

Between the transmissions of OFDM frames\footnote{It may be conditioned to when a temperature variation of the HPLED is detected or by extrapolation of an operating threshold time.}, the DPD block disconnects from data transmission mode and enters the compensation mode. 
Considering that $\bm{I}_{\rm dpd}$ and $\bm{I}$ are the output and input currents of the DPD, respectively, the compensation mode in the DPD consists in estimating a compensation for a polynomial function in $\bm{I}_{\rm dpd}(\bm{I})$ in order to make linear the relationship between $\bm{V}_{\rm tia}(\bm{I})$. For this, the DPD method presented in \cite{2009elgala_dpd_ofdm} was adapted\footnote{The DPD original technique was developed in relation to $I_{\rm led} (V_{\rm led})$.}. Here is a description of each step:
\begin{algorithm}\caption{Digital Pre-Distortion (DPD) Procedure}
	\begin{enumerate}
	\item A sequence of $J$ equally spaced current levels and within the dynamic range of the HPLED is generated:
	\begin{equation}
	\bm{I} [j] = I_l  + \frac{{(I_u  - I_l )j}}{{J-1}}; \quad j = 0, \ldots ,J-1.
	\end{equation}
	with $\bm{I}_{\rm dpd} = \bm{I} + I_{\rm bias,0}$, and $I_{\rm bias,0}$ the initial DC bias current.
	
	\item For each current $\bm{I}_{\rm dpd}$, estimates the average of the samples of $\bm{V}_{\rm tia}$ which has been converted by the ADC.
	
	\item Perform a polynomial fitting of previous experimental points:
	\begin{equation}
	\bm{I}_{dpd} = a \bm{V}_{\rm tia} \circ \bm{V}_{\rm tia} + b\bm{V}_{\rm tia} + c.
	\label{eq:idpd_vtia}
	\end{equation}
	
	\item Generate the vector, also equally spaced:
	\begin{equation}
	\bm{\dot V}_{\rm tia} [j] = \min(\bm{V}_{\rm tia})  + \frac{{\left(\max(\bm{V}_{\rm tia})  - \min(\bm{V}_{\rm tia}) \right) j}}{{J-1}},
	\end{equation}
	with $j = 0, \ldots ,J-1$.
	
	\item Determine $\bm{\dot I}_{\rm dpd}(\bm{\dot V}_{\rm tia})$ by applying \eqref{eq:idpd_vtia}.
	\item Perform a second order polynomial fitting for $\bm{\dot I}_{\rm dpd}(\bm{I})$.
\end{enumerate}
\end{algorithm}

\begin{figure}[!htbp]
\centering\includegraphics[width=.69\textwidth]{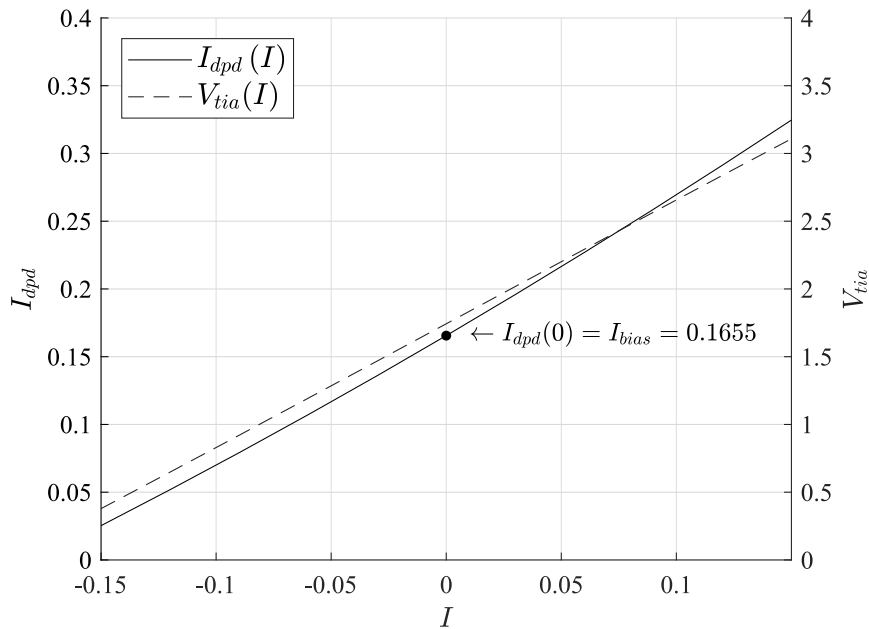}
\caption{Example of DPD block output and TIA output as a function of the input signal $\bm{I}$.}
	\label{fig:plot_dpd} 
\end{figure}

Thus, the adjustment parameters of $\bm{\dot I}_{\rm dpd}(\bm{I})$ are applied to the function $\bm{I}_{\rm dpd}(\bm{I})$ 
when in transmission mode. For instance, Fig. \ref{fig:plot_dpd} is constructed with the output of the DPD and the output of the TIA as a function of the input signal $\bm{I}$. In this case, $I_{\rm bias,0}=175$ mA, $I_u=-I_l=150$ mA, and the parameters of the Table \ref{tab:fit_parameters} are considered for the temperature of 50$\degree$C. Notice that by keeping the dynamic current range of the HPLED fixed, the bias current changes, however, it is easily calculated by $\bm{I}_{\rm dpd}(\bm{I}=0)$. Finally, the signal after DPD is converted from digital to analog (DAC)
 electrical current and finally coupled to the HPLED.

\subsection{GE block and Pre-Equalization} \label{subsec:pre_eq}
The analysis of the GE block is divided into two situations, considering both flat and non-flat electric gain $\bm{G}_E$ response. Hence, the  pre-equalization scheme is implemented for the non-flat electric gain response case.
\subsubsection{GE block in Electrical Gain Flat}\label{subsubsec:ge_flat}
Although the flat electric gain $\bm{G}_E$ is not purely real-valued, it was initially considered for the ceteris paribus type analyzes. An example of such approach is developed in the Subsection \ref{subsec:simu_dpd}. Hence, since the mean value of the signal purely real $\bm{x}$ is null, i.e., the DC level is null, its variance can be determined by applying the Parseval theorem:
\begin{equation}
\sigma_x^2=\mathbb{E}[\bm{x}[i]^2]=\mathbb{E}[|\bm{x}[i]|^2] = \frac{1}{N} \sum\limits_{i = 0}^{N - 1} {|\bm{x}[i]|^2 } = \frac{1}{N^2} \sum\limits_{i = 0}^{N - 1} {|\bm{X}[i]|^2 }.
\end{equation} 
Moreover, considering the Hermitian symmetry of $\bm{X}$, the element $\bm{X}[0]=0$ responsible for the DC level, and the concept of the operator expectation, one can write the variance:
\begin{equation}
\begin{array}{l}
\sigma_x^2=\frac{2}{N^2} \sum\limits_{i = 1}^{\frac{N}{2} - 1} {|\bm{X}[i]|^2 } 
=\frac{N-2}{N^2} \,\, \mathbb{E}[|\bm{X}[i]|^2].
\end{array}
\end{equation}
Considering eq. \eqref{eq:gain_multiplication}, and that $\bm{X}_H$ and $\bm{G}$ are independent random variables:
\begin{equation}
\sigma_x^2= \frac{N-2}{N^2}  {\mathbb{E}[|\bm{X_H} \circ \bm{G}|^2] } =\frac{N-2}{N^2} \mathbb{E}[|\bm{X_H}[i]|^2] \mathbb{E} [|\bm{G}[i]|^2];
\label{eq:sig2_3}
\end{equation} 
where $\mathbb{E}[|\bm{X_H}[i]|^2]=\frac{2}{3}(M-1)$ 
is the mean power of the modulated QAM symbol with $M$ order.

Taking into account 
the flat channel, the value of the elements of the vector $\bm{G}$ are constants, i.e., $\mathbb{E} [|\bm{G}[i]|^2]=|G|^2$. Applying this result in \eqref{eq:sig2_3}, the module of the $G$ elements are determined by:
\begin{equation}
|G|_{\rm flat}=\sqrt{\frac{3\sigma_x^2 N^2}{2(N-2)(M-1)}}.
\label{eq:g_flat}
\end{equation}

\subsubsection{GE block in Non-Flat Electrical Gain}\label{subsubsec:ge_non_flat}
In the non-flat electrical gain response, the luminous feedback signal is used to equalize the received OFDM symbol due to frequency selective attenuation. Considering the scheme of Fig. \ref{fig:proposed_scheme}, the pre-equalization process is summary described as follows. 
\begin{algorithm}\caption{Pre-equalization Frequency Domain Procedure}
\begin{enumerate}
	\item Transmitting the symbols of the OFDM subcarriers deploying constant elements of vector $\bm{G}$ given by \eqref{eq:g_flat};
	\item capturing the feedback signal by the ADC; 
	\item  Estimating the symbols received through the FFT block;
	\item Determine new $\bm{G}$ gains to be applied to the subcarriers: \\
	- power composition of the signal $\bm{u}$ remains within the dynamic range of HPLED, \eqref{eq:Fator_Clipping}.
\end{enumerate}
\end{algorithm}

The GE block uses an estimator by zero forcing (ZF). In this way, the gain vector of the subcarriers is determined by:
\begin{equation}
\bm{G}=\alpha \oslash \bm{Y}_{FB},
\label{eq:zf}
\end{equation}
where $\alpha$ is a scaling factor, $\oslash$ is the element-wise division and $\bm{Y}_{FB}$ is the output of the FFT block implemented in the VLC transmitter feedback loop. Considering \eqref{eq:zf} in \eqref{eq:sig2_3}, $\alpha$ can be determined as:
\begin{equation}
\alpha= \sqrt{ \frac{3\sigma_x^2 N^2}{2(N-2)(M-1)}\cdot\frac{1}{\overline {|\bm{Y}_{FB}|^2}}},
\label{eq:alpha_ge_non_flat}
\end{equation}
where $\overline {|\bm{Y}_{FB}|^2}$ is the  mean squared of the module of the elements of the vector $\bm{Y}_{FB}$.

\subsection{DCO-OFDM VLC receiver} \label{subsec:dco_ofdm_vlc_rx}
At the receiver, after FFT, the M-QAM symbols are mapped and sorted in an inverse process to that described by \eqref{eq:XH}. 
Then the post-equalization is applied also by ZF using:
\begin{equation}
\bm{Y}_D=\bm{Y}_d\, \circ\, \bm{C}_{\rm post}
\end{equation}
 where the estimates $\bm{C}_{\rm post}$ 
 is obtained by element-wise division of pilot symbols by $\bm{X}_D$ vector, $\bm{C}_{\rm post}=\bm{Y}_D \oslash \bm{X}_D$.
In the case with Pre-Eq, {\it i.e.} pre-post-equalization (PP-Eq), the vector $\bm{C}_{\rm post}$ is determined after the estimation of $\bm{G}$ and the transmission of a new OFDM symbol already pre-equalized.
Finally,  the demodulation is performed to estimate the received data.

\section{Numerical and Experimental Results}\label{sec:results}
In this section, we have demonstrated the effectiveness and efficiency of the proposed method through numerical simulation and experimental setup analyses. The adopted system parameter values are presented in Table \ref{tab:parameters}. The transmitter HPLED and the PD were perfectly aligned, having angles $\phi_{\rm rx}=\theta_{\rm rx}=0^o$. The feedback PD has been inclined to receive part of the light emitted by the HPLED, i.e., $\phi_{\rm fb}=45^o$ and $\theta_{\rm rx}=0^o$. 
With the clipping factor $\gamma=5$, the variance can be determined by \eqref{eq:Fator_Clipping}, resulting in a standard deviation $\sigma_x=0.03$.
Using \eqref{eq:omega_rx} and \eqref{eq:omega_fb}, the optical gain of the channels resulted $\Omega_{\rm rx}=5.968 \cdot 10^{-8}$ and $\Omega_{\rm fb}=1.004 \cdot 10^{-3}$.
 
\begin{table}[!htbp]
	\centering
	\caption{Adopted parameters values.}
	\begin{tabular}{l|l|l}
		\hline	
		\multicolumn{1}{c|}{\bf HPLED}&\multicolumn{1}{c|}{\bf PDs*}&\multicolumn{1}{c}{\bf OFDM}\\
		\hline\hline
		$\bm{r}_{\rm led} = [2,2,3]$		& $\bm{r}_{\rm rx} = [2,2,1]$		& $B_{\rm OFDM}=5$ MHz \\  
		$\bm{n}_{\rm led} = [0,0,-1]$		& $\bm{n}_{\rm rx} = [0,0,1]$		& $N=1024$ \\
		$n_L=0.5$    						& $\bm{r}_{\rm fb} = [1.98,2,2.98]$	& $\gamma=5$\\
		$I_{\rm bias,0}=175$ mA 				& $\bm{n}_{\rm fb} = [1/\sqrt{2},0,1/\sqrt{2}]$	&  \\
		$I_u=150$  mA 						& $A_{\rm pd}=$ 1 mm$^2$			& \\
		$I_l=-150$ mA 						& $R_{\rm pd}=$ 0.54 A/W			& \\
		\hline
		\multicolumn{3}{c}{{\it *Distance between HPLED and PD}:\quad $d_{\rm tx-rx}\in[40; \, 110]$ cm}\\
			\hline
	\end{tabular}%
	\label{tab:parameters}%
\end{table}%

\subsection{Simulation of the DPD in Flat Electrical Gain} \label{subsec:simu_dpd}

\begin{figure}[htbp!] 
\centering\includegraphics[width=.99\textwidth]{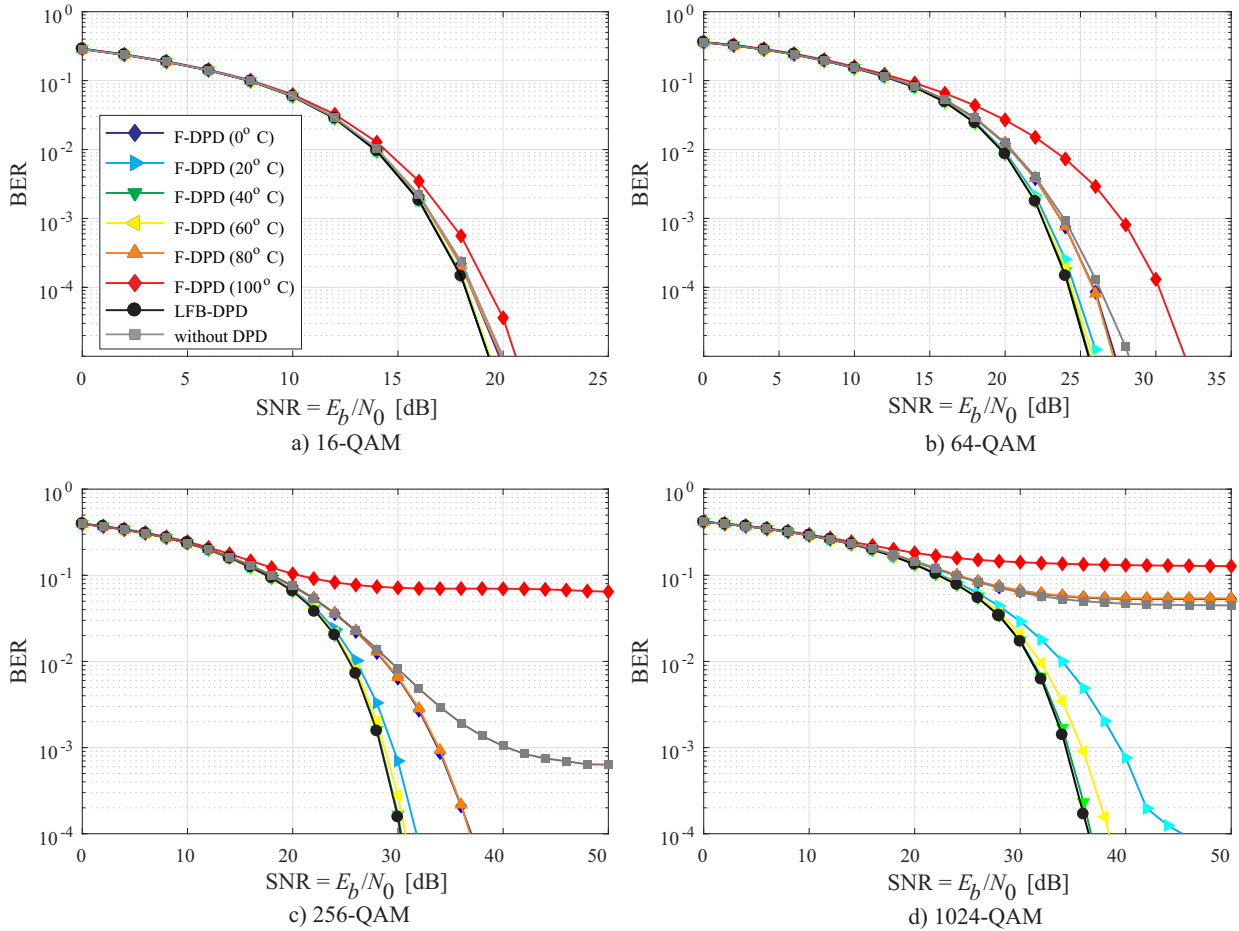}
\caption{BER simulation for fixed DPD (F-DPD) with different temperatures in the HPLED, for DPD by luminous feedback (LFB-DPD) and for without DPD. All for different modulation orders in flat channel.}
	\label{fig:ber_dpd} 
\end{figure}

Considering Monte Carlo simulations (MCS), Fig. \ref{fig:ber_dpd} depicts the bit error rate (BER) {\it versus} signal-to-noise ratio (SNR) results for the system with fixed pre-distortion (F-DPD) calibrated for fixed temperature of 50$\degree$C.  
The simulations were performed for different temperatures emulated by the HPLED model of Table \ref{tab:fit_parameters}. 
Specifically temperatures of 0, 20, 40, 60, 80 and 100$\degree$C have been considered in the analyses of this section. 
In the same graphs, the results of the proposed digital pre-distortion by luminous feedback (LFB-DPD) and the system without DPD (W-DPD) are presented.

In Fig. \ref{fig:ber_dpd}  it is possible to verify that for the F-DPD and for the system without DPD, a greater effect in the degradation of the performance occurs for larger orders of modulation. This can be attributed to the fact that the Euclidean distance between the QAM symbols is much smaller, being more susceptible to the detection errors generated by the HPLED nonlinearity effects. 
In some cases, the result becomes so poor that several BER curves stagnated asymptotically close to the range of $0.05 \leq BER \leq 0.15$ (BER floor). 
Besides, in some situations of F-DPD, the performance is worse than the system without pre-distortion, {\it i.e.}, in such cases it is better to perform without pre-distortion scheme. 
These results corroborate the work proposal, in which it is more advantageous to have the pre-distortion by the luminous feedback signal. In the range of modulation orders analyzed, LFB-DPD was the scheme that presented the best performance.

\subsection{Proposed Experimental Setup}\label{subsubsec:experimental_setup}

\begin{figure}[!htbp]
\centering\includegraphics[width=.67\textwidth]{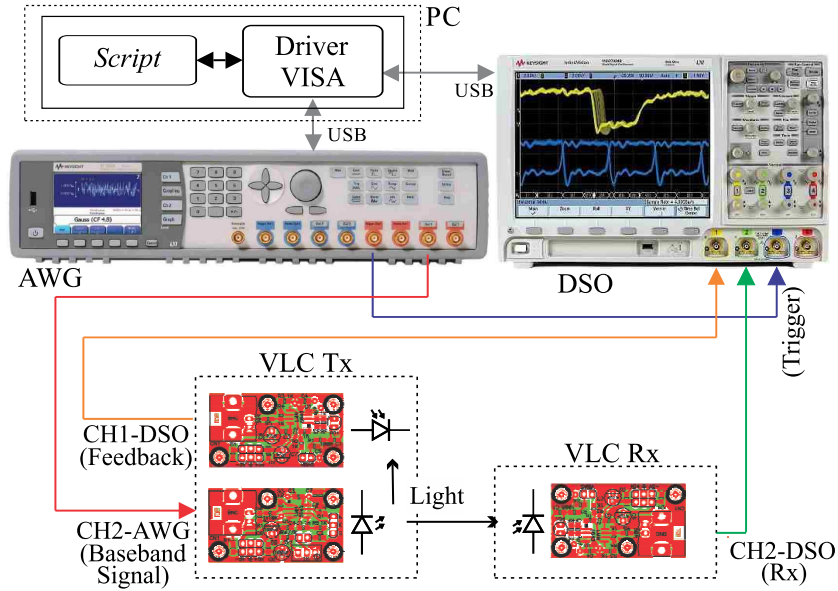}
	\caption{Experimental arrangement for the validation of the proposed architecture.}
	\label{fig:experimental_scheme} 
\end{figure}

The experimental setup for validation of the  digital pre-distortion and pre-equalization schemes in the real physical system (RPS) OFDM-VLC system is sketched in Fig. \ref{fig:experimental_scheme}. The arrangement was implemented in order to verify the proposed schemes using a playback-type approach, {\it i.e.}, the signals to be generated by the arbitrary wave generator (AWG) and the signals captured by the digital storage oscilloscope (DSO) are processed off-line in a personal computer (PC). 

\begin{figure}[!htbp] 
\centering\includegraphics[width=.47\textwidth]{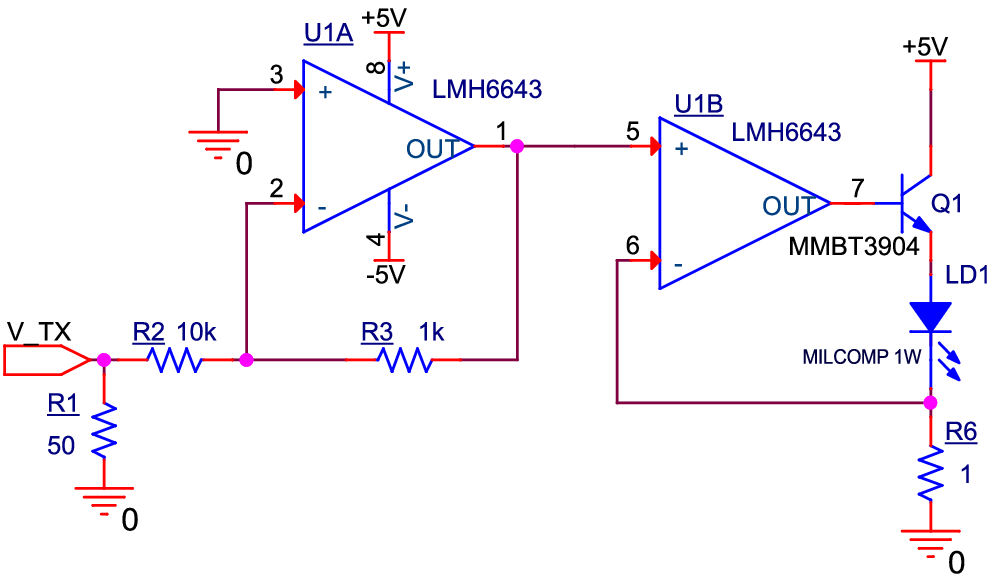}
\hspace{.8cm}
\includegraphics[width=.42\textwidth]{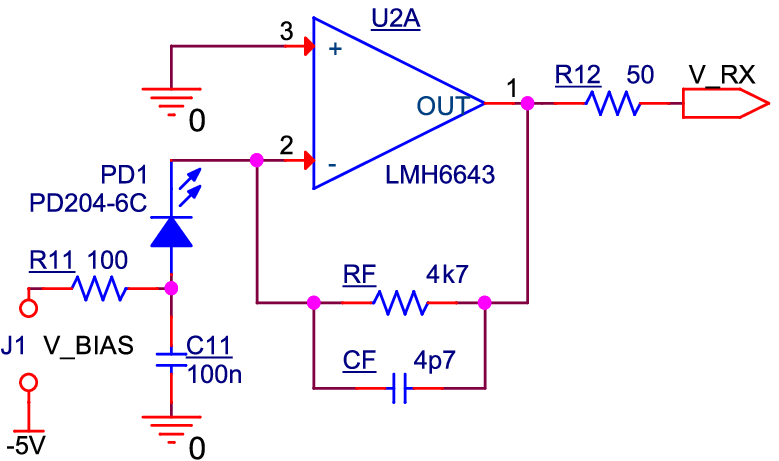}\\ 
a) \hspace{40mm} b)
\caption{Proposed circuit for HPLED driver in (a) and TIA circuit in (b).}
	\label{fig:driver_tia_circuit} 
\end{figure}

Thus, for the light interface, three circuits were implemented, one light transmitter and two light receivers.
Fig. \ref{fig:driver_tia_circuit}a shows the electronic schematic of the transmitter that has the function of modulating the current signal on the HPLED.
Fig. \ref{fig:driver_tia_circuit}b shows the TIA circuit\footnote{The J1 jumper is kept closed to the polarization of the PD.} used in the feedback receiver and the remote receiver. 
The TIA project was based on \cite{ramus2009transimpedance} and has bandwidth $B$ = 10.01 MHz.
As presented in Section \ref{sec:intro}, the reverse polarization has the purpose of reducing the capacitance of the PD and consequently increasing its speed and operating frequency band.

\begin{figure}[!htbp]
\centering\includegraphics[width=.67\textwidth]{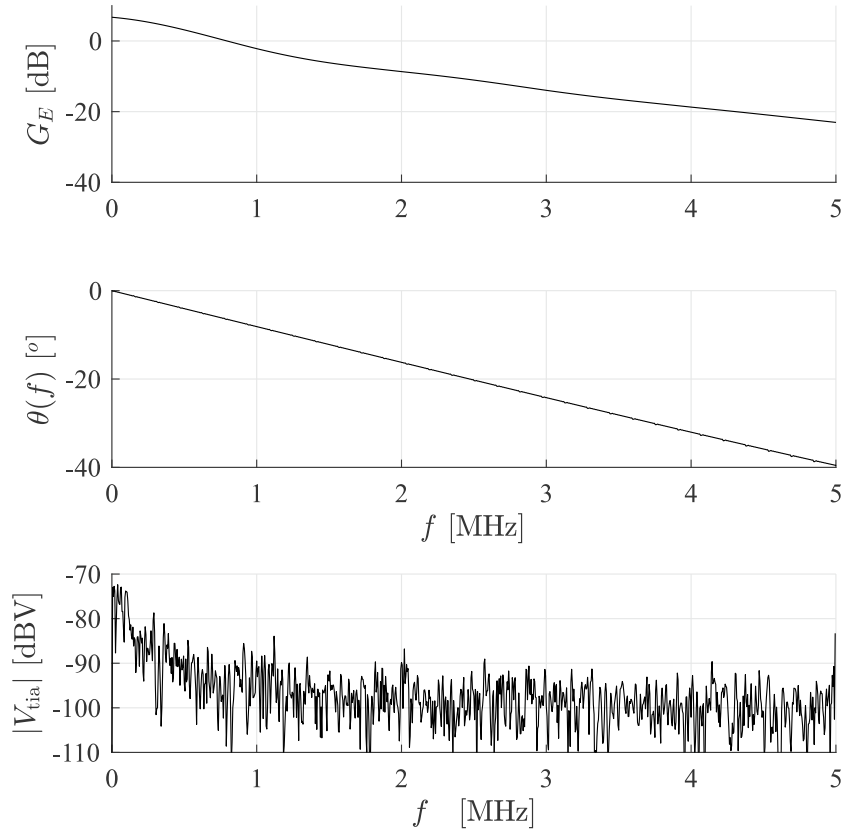}
\caption{Frequency response of the electric gain and noise after TIA.}
	\label{fig:resp_freq_noise} 
\end{figure}

As the first experimental verification, it was obtained the frequency response of the electric gain $G_E$. Fig. \ref{fig:resp_freq_noise} shows the behavior of the amplitude and phase of the system with the 1W Multicomp HPLED. The GE gain was extracted by generating an electric frequency sweep signal applied to the HPLED driver and extracting the spectral magnitude of the electric signal obtained after the TIA. Fig. \ref{fig:resp_freq_noise}  also illustrates the noise captured after the TIA. It was obtained in a similar way to the extraction of the GE; however, applying a DC signal of $I_{\rm led} = I_{\rm bias,0} = 0.175$ mA in the HPLED. It can be seen from Fig. \ref{fig:resp_freq_noise}  that the noise in the range of a few hundred kHz is about 25 dB higher than the noise floor.

\subsection{Simulation with LFB-DPD in Non-Flat Electrical Gain}\label{subsec:simu_dpd_non_flat}

\begin{figure}[!htbp]
\centering\includegraphics[width=.67\textwidth]{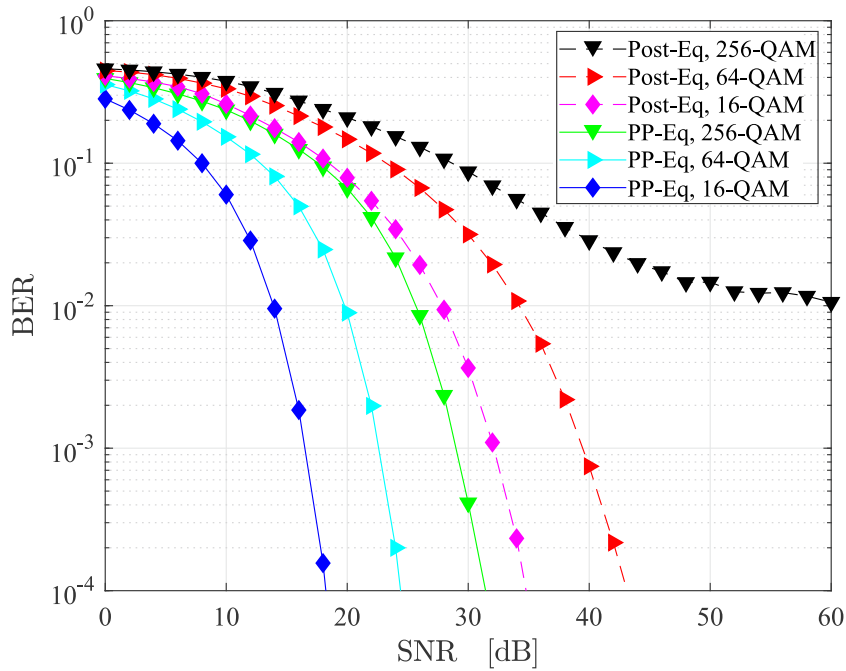}
\caption{BER for proposed scheme with PP-Eq and Post-Eq in Non-Flat Electrical Gain.}
	\label{fig:ber_pre_pos_equalizer} 
\end{figure}

\begin{figure}[!htbp]
\centering\includegraphics[width=.3\textwidth]{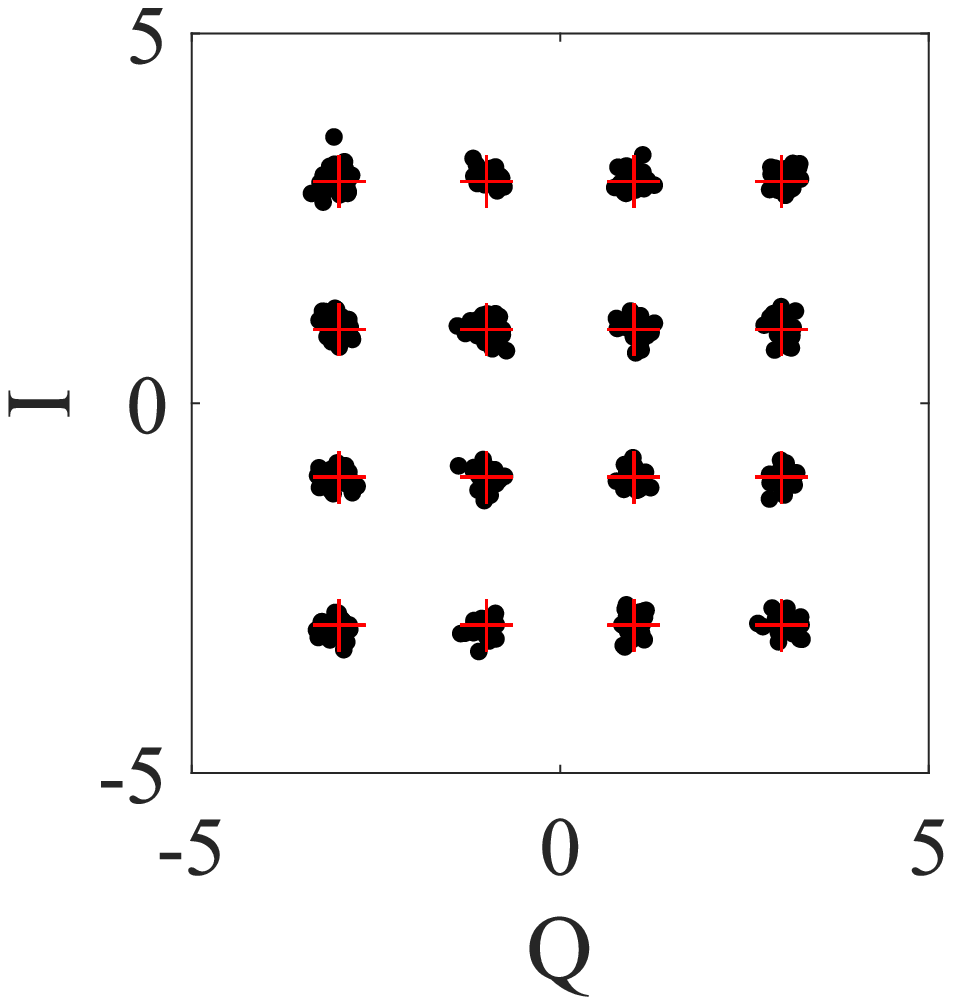}\hspace{18mm}
\includegraphics[width=.3\textwidth]{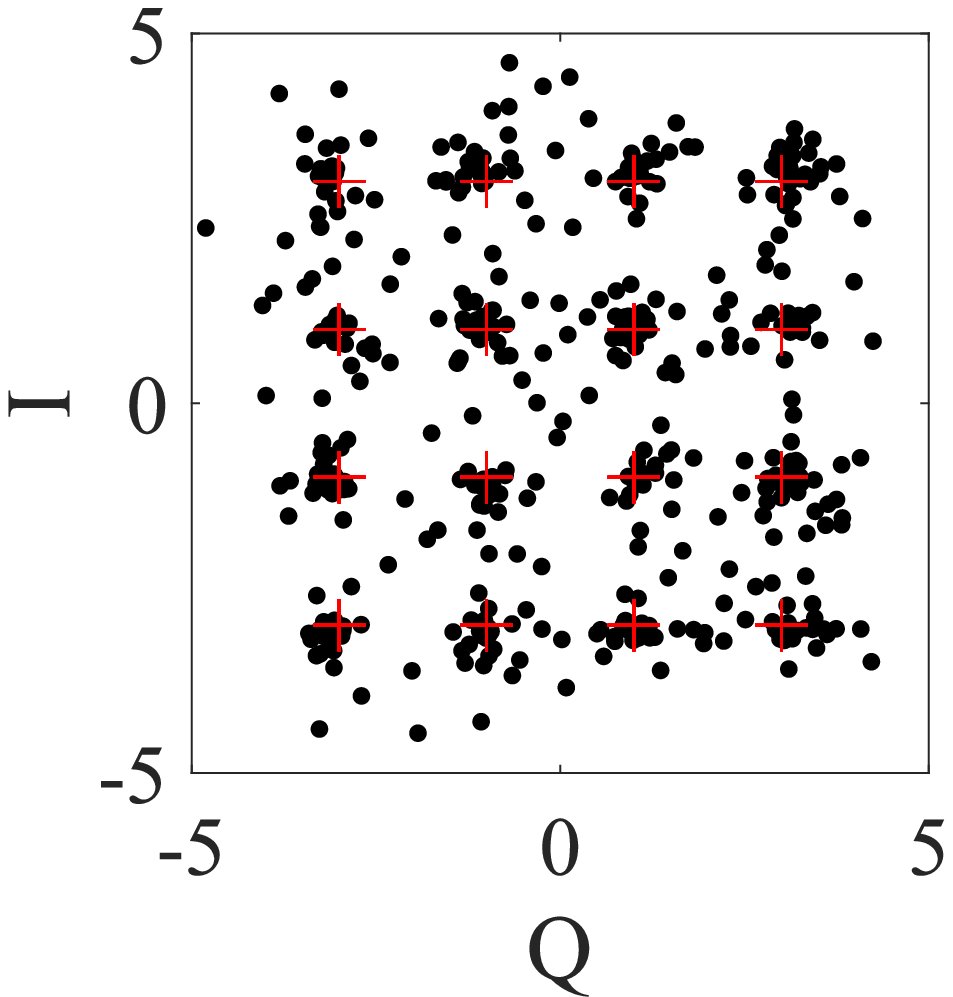}\\
a) Pre-post-equalization (PP-Eq) \hspace{18mm} b) Post-equalization (Post-Eq)
\caption{16-QAM constellation scatter plot at SNR = 25 dB in the OFDM frame.}
	\label{fig:const_pre_pos} 
\end{figure}

From the result of the $G_E$ obtained in subsection \ref{subsubsec:experimental_setup}, the system simulation for the non-flat channel was implemented. Fig. \ref{fig:ber_pre_pos_equalizer} presents the simulation results for different modulation orders (16-QAM to 256-QAM) for the proposed system with pre-post-equalization (PP-Eq) and with only post-equalization (Post-Eq). One can check the performance improvement obtainec with PP-Eq system compared to the Post-Eq. Indeed, Post-Eq performance worsens in such a way  that under 256-QAM modulation the BER performance stagnates asymptotically, presenting a BER floor of $\sim 10^{-2}$. 
This is due to the fact that the higher frequency subcarriers suffer from higher attenuation, therefore, resulting in a lower SNR in case of the Post-Eq strategy.
Notice that applying the pre-equalization (Pre-Eq), such effect is mitigated, offering a compensation by transmitting more power in the attenuated subcarriers, in order to maintain a suitable average SNR across all subcarriers. Fig. \ref{fig:const_pre_pos} illustrates this difference by displaying the 16-QAM I-Q constellation scatter plot, considering PP-Eq and Post-Eq at the same SNR of $25$ dB in the OFDM frame. It is possible to verify the greater spreading of the symbols in the Post-Eq, and consequently, the greater symbol error occurrences.

\subsection{Validation in the RPS of the complete system}

For the validation of the proposal by means of the experimental arrangement presented in Subsection \ref{subsubsec:experimental_setup}, basically, the same parameters presented in the introductory part of this section were considered. Changes include: a) the inversion of the $z$ coordinates with $y$ in order to facilitate the experiment mounting on a table; b) variation of the distance between the HPLED transmitter and the VLC receiver photodiode; c) the last change allows to control the SNR aiming at  evaluating the BER, {\it i.e.}, the closer, the greater the SNR. However, it has kept fixed the position of the transmitter HPLED and the feedback photodiode, while hold controlled the ambient temperature at 25$\degree$C.

In accordance with the discussion of the frequency response in Fig. \ref{fig:resp_freq_noise}, the first hundred OFDM subcarriers were deactivated in the implementated circuit due to the higher noise power verified across this frequency range. 
If enabled, these subcarriers would degrade overall system performance.
Thus, Eq. \ref{eq:alpha_ge_non_flat} should consider the $N_s$ suppressed subcarriers. In this case, the parameter $\alpha$ was reformulated as:
\begin{equation}
\alpha'= \sqrt{ \frac{3\sigma_x^2 N^2}{2(N-2N_s)(M-1)}\cdot\frac{1}{\overline {|\bm{Y}_{FB}|^2}}}.
\label{eq:alpha_ge_non_flat_supressed}
\end{equation}

\begin{figure}[!htbp]
\centering\includegraphics[width=.49\textwidth]{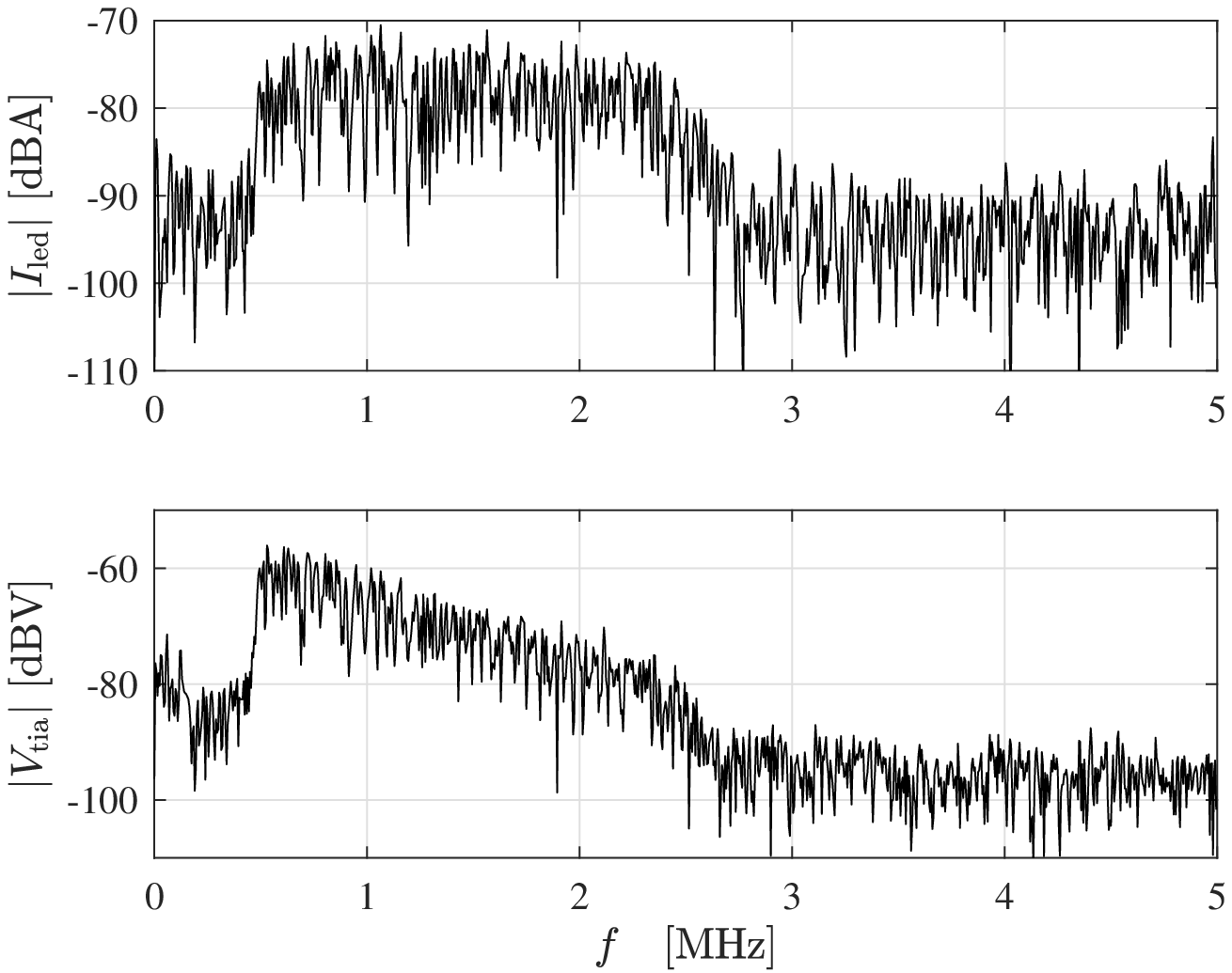}
\,\includegraphics[width=.49\textwidth]{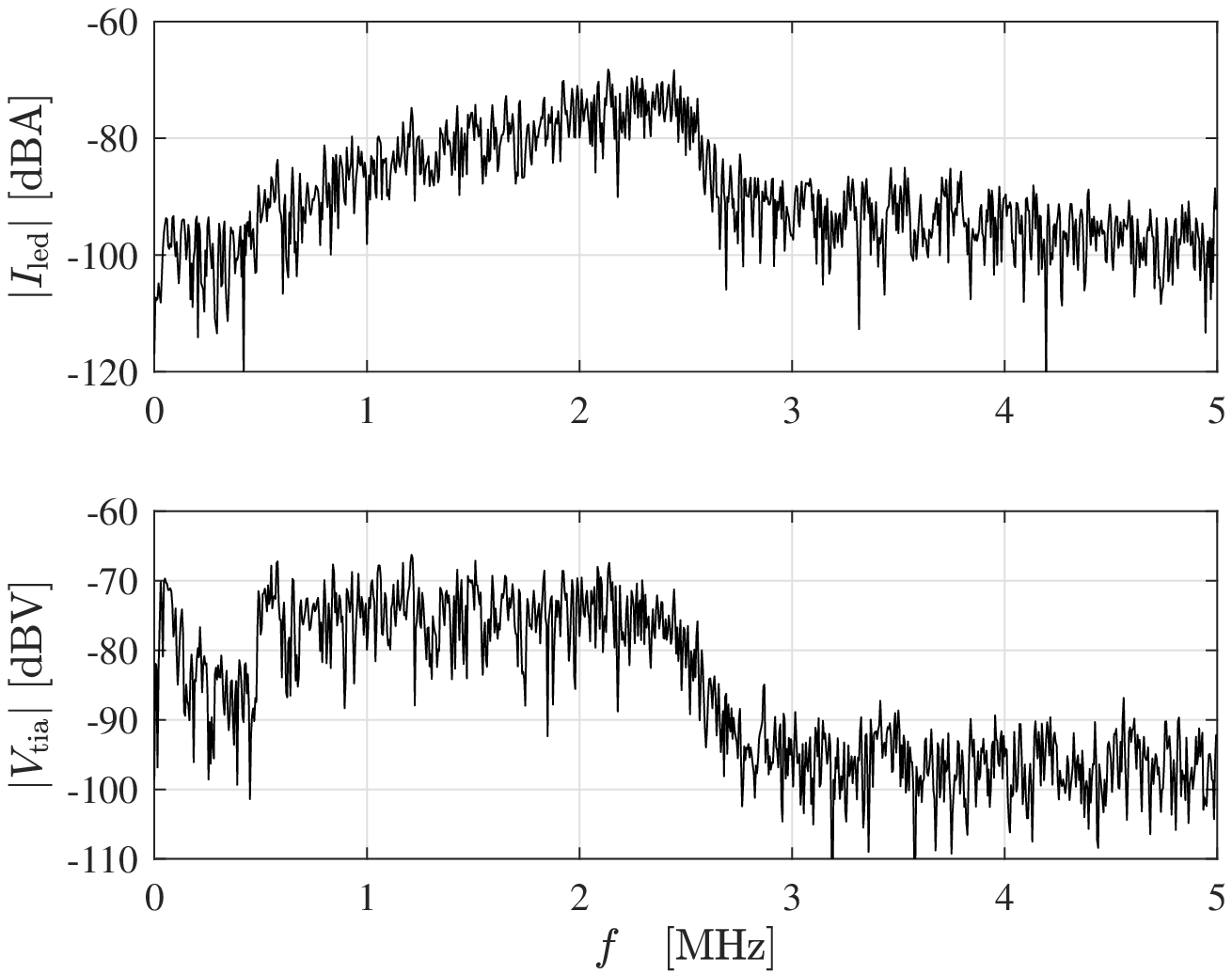}\\
\hspace{-10mm} a) subcarriers with equal powers \hspace{25mm}  b) pre-equalized powers
\caption{Signal spectrum transmitted (upper) and received (lower) for the active subcarriers with the same power in a) and with the Pre-Eq strategy in b).}
	\label{fig:espectro_ofdm_sfr} 
\end{figure}

Fig. \ref{fig:espectro_ofdm_sfr} depicts the 16-QAM OFDM signal spectrum applied in the HPLED and the captured signal after the TIA of the receiver at distance of $d_{\rm tx-rx}=40$ cm, for the condition in which the active subcarriers are transmitted with the same power and under pre-equalization (Pre-Eq).
In the case of the Pre-Eq, it is possible to observe the higher power transmitted in the higher frequency subcarriers as a way of compensating the respective larger attenuations. The last graph of Fig \ref{fig:espectro_ofdm_sfr} illustrates the maintenance of the same power ceiling of the subcarriers at the receiver side due to the pre-equalization effect. Also in this graph, it is possible to verify that the noise reaches the same order of magnitude of the OFDM signal received in the range of disabled subcarriers.

\begin{figure}[!hbtp] 
\centering\includegraphics[width=.69\textwidth]{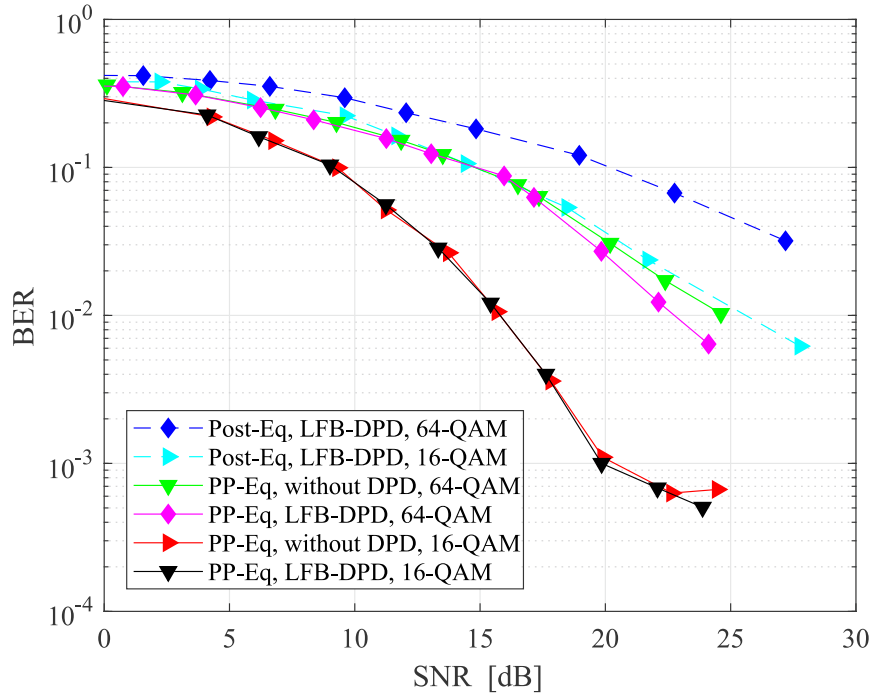}
	\caption{BER by the experimental setup.}
	\label{fig:ber_sfr} 
\end{figure}

By varying the distance between the transmitter HPLED and the receiver photodiode within the range of $40-110$ cm it was possible to obtain the Fig. \ref{fig:ber_sfr} with the BER performance of the real system\footnote{The minimum and maximum range can be changed by implementing an automatic gain control (AGC) on the VLC receiver circuit.}.
The capture of the BER for distances smaller than 40 cm was avoided due to saturation of the TIA of the VLC receiver under these conditions.
This fact limited the evaluation of the system to maximum SNR in the VLC receiver of about 24 dB  and 27.5 dB under PP-Eq and Post-Eq schemes, respectively.
Overall, the great improvement in PP-Eq performance is confirmed comparing with the BER performance attained with the Post-Eq OFDM-VLC strategy.
Particularly in larger SNRs, the BER improvement was also confirmed for the PP-Eq LFB-DPD configuration operating under high order modulation (64-QAM), when compared to the equivalent 64-QAM without digital pre-distortion (DPD). 
Moreover, for the PP-Eq with 16-QAM case, the improvement was marginal in the covered SNR range. This way, the RPS setup confirms the results obtained by simulation in Subsections \ref{subsec:simu_dpd} and \ref{subsec:simu_dpd_non_flat}.
Besides, Fig. \ref{fig:ber_sfr} reveals a BER floor in the performance of PP-Eq under 16-QAM configuration; one can verify a BER floor around $5\cdot 10^{-4}$ for SNR $\geq 20$ dB. Because it is close to the saturation point of the TIA amplifier, this BER floor can be attributed to the backoff effect of the operational amplifier and to the noise in the active subcarrier range that is not exactly AWGN, to the oscilloscope DAC quantization error, to the noise generated by the clipping and to the possible synchronization errors. 
Considering that in this experimental arrangement we have used general-purpose electronic components, the performance can be improved substantially in high SNR regime using HPLEDs and photodiodes with better characteristics for VLC applications such as devices with lower intrinsic capacitances, lower noise and greater sensitivity of the visible light range of the photodetectors. 

\section{Conclusion}\label{sec:conc}
In this work, by means of an experimental setup, the nonlinear behavior of the optical power emitted by an HPLED was characterized as a function of the current at different temperatures at its semiconductor junction. 

The proposed digital pre-distortion with light feedback (LFB-DPD) scheme implemented at the transmitter side, allowed improve the system performance by mitigating the nonlinearity effect.
The numerical simulation results demonstrated that the pre-distortion with fixed parameters (F-DPD) presents even worse performances than the system without DPD in some cases. 
This reinforces the necessity of the proposed light feedback scheme implemented in the transmitting device.

This scheme also enabled the implementation of a pre-equalization (Pre-Eq) on the OFDM subcarriers. It was confirmed that pre-post-equalization (PP-Eq) promoted a great improvement over BER performance when compared to the post-equalization (Post-Eq) scheme, since it enables the receiver to maintain an average SNR across the OFDM subcarriers. In this way, an experimental arrangement was developed aiming at extracting the system parameters for computational simulations purpose and also for the validation of the proposed system. 
Beyond validating the DPD model, the physical implementation of the complete OFDM-VLC system, also confirmed a much better overall system performance with pre-post-equalization (PP-Eq) strategy. 
 
Finally, although it has not been analyzed in this work, the variation of luminosity and nonlinearity due to the ageing of the HPLED, the proposed architecture is able to correct the component ageing effects through small adaptations.

\section*{Funding}
This work has been partially supported by the National Council for Scientific and Technological Development (CNPq) of Brazil under Grants 304066/2015-0; by the Londrina State University (UEL) and the Parana State Government. All the agencies are gratefully acknowledged.



\end{document}